\documentstyle[12pt,epsf]{article}
\newcommand{\nn}{\nonumber}
\newcommand{\be}{\begin{equation}}
\newcommand{\ee}{\end{equation}}
\newcommand{\bea}{\begin{eqnarray}}
\newcommand{\eea}{\end{eqnarray}}
\newcommand{\bean}{\begin{eqnarray*}}
\newcommand{\eean}{\end{eqnarray*}}
\newcommand{\ag}{\alpha}
\newcommand{\ay}{\alpha_y}
\newcommand{\tl}{t_{\Lambda}}
\newcommand{\etab}{\bar{\eta}}

\begin{document}
\vspace{-1cm}
\noindent
\begin{flushright}
KANAZAWA-00-04\\
KUNS-1666
\end{flushright}
\vspace{10mm}
\begin{center}
{\Large \bf 
Softly Broken Supersymmetric Gauge-Higgs-Yukawa 
\vspace{2mm}\\
Theories as Renormalizable Composite Models
}
\vspace*{15mm}\\
Tatsuo Kobayashi$^a$
\footnote{E-mail: kobayashi@gauge.scphys.kyoto-u.ac.jp}
and Haruhiko Terao$^b$
\footnote{E-mail: terao@hep.s.kanazawa-u.ac.jp}
\vspace*{5mm}\\
$^a$Department of Physics, Kyoto University\\ 
Kyoto 606-8502, Japan 
\vspace{2mm}\\
$^b$Institute for Theoretical Physics, Kanazawa University\\
Kanazawa 920-1192, Japan
\end{center}
\vspace*{10mm}
\begin{abstract}
We examine the (softly broken) supersymmetric gauge-Higgs-Yukawa
theories satisfying the compositeness conditions at a certain scale.
In these theories the Higgs superfields can be regarded as the chiral
composite fields. It is found that there are the fundamental theories, 
which contain the dimension 5 interactions and the hard SUSY breaking 
in perturbation, 
turn to be renormalizable and also softly broken 
theories in the nonperturbative framework. 
The soft SUSY breaking parameters as well as the Yukawa coupling
in the corresponding gauge-Higgs-Yukawa theories are restricted by the 
renormalization group invariant relations.
\end{abstract}
\vspace*{20mm}
\noindent
PACS numbers: 11.30.Pb, 11.10.Hi, 11.10.Gh, 12.60.Fr
\\
Keywords: composite higgs, renormalizability, renormalization group, coupling
reduction, soft supersymmetry breaking

\newpage
\pagestyle{plain}
\pagenumbering{arabic}

\section{Introduction}

Supersymmetric extension of the Standard Model (SM) brings much more
parameters after supersymmetry (SUSY) breaking than in the SM. 
Most of them are soft SUSY breaking parameters.
It is clearly attractive to reduce these plenty of free parameters
by some model independent mechanisms. 
Especially it is an important problem to clarify the SUSY breaking
origins leading the viable phenomenological models.
So far scenarios based on the gravity-mediated and the gauge-mediated
supersymmetry breaking mechanisms have been mostly investigated.
Recently, however, there is a great interest in the anomaly-mediated
supersymmetry breaking \cite{anomaly}, 
which restricts the soft SUSY breaking parameters to the 
renormalization group (RG) invariant
relations \cite{hs,jjp,jj1,kkk}. 
Therefore this SUSY breaking mechanism leads to highly 
predictable phenomenological models. 
However there is the problem that the slepton masses become tachyonic.
Therefore, another source of SUSY breaking has to be introduced
in order to resolve this problem \cite{anomaly} \footnote{
See also Ref.\cite{dterm}, where other contributions 
are added with respecting the anomaly mediation RG trajectory.}.

In this paper we propose another possibility to reduce the
number of the free parameters in supersymmetric theories.
We consider the supersymmetric gauge-Higgs-Yukawa theories, whose
Higgs chiral superfields are composites of the fundamental 
matter chiral superfields.
The scenarios beyond the SM with composite Higgs particles,
e.g. Technicolor models, Top condensation models and so on, have been
intensively examined before.
In the context of supersymmetry also, the dynamics of composite particles 
in the Nambu-Jona-Lasinio (NJL) models has been 
studied \cite{susynjl,susynjl2}.
The Higgs particles are given as composites of fermions in these models.
The models considered in this paper are a kind of supersymmetric extension
of the Top condensation models \cite{composite}.
However the Higgs particles in our models are composites of scalars.
In practice we seek for the special solutions of 1-loop RG equations
for the (softly broken) gauge-Higgs-Yukawa theories which satisfy
the compositeness conditions at a certain scale \cite{composite}.

The RG equations for the non-supersymmetric gauge-Higgs-Yukawa theories
have been examined in the leading order of a modified $1/N_c$ expansion
in Ref.~\cite{hkkn}.
It was found that the nontrivial continuum limits exist for a certain
class of theories.
Moreover the gauged NJL models become equivalent to the
nontrivial gauge-Higgs-Yukawa theories in the infinite cutoff limit.
This proves the renormalizability of the gauged NJL models in four
dimensions.
Later this renormalizability was also confirmed directly by applying the
Exact RG to the gauged NJL models \cite{kt}.
The Yukawa coupling of the theory equivalent to the renormalizable
gauged NJL model is restricted to the Pendelton-Ross fixed point 
\cite{pr}. The fixed point solution has been also obtained by the
coupling reduction \cite{reduction,kubo}.
The reduction of couplings is a natural consequence,
since the gauged NJL models contain less parameters than the
gauge-Higgs-Yukawa theories. 

If we consider the supersymmetric extension of the gauge-Higgs-Yukawa
theories, similar mechanism is found to work as well. 
In this case the interactions of the fundamental theory are given by
dimension 5 operators. It will be shown that these theories also can be
renormalizable similarly to the gauged NJL models. 
Moreover, the SUSY breaking terms are introduced to these theories. 
Then it is found
that a kind of hard SUSY breaking terms can be introduced in the dimension
5 interaction models, while the equivalent gauge-Higgs-Yukawa theories
are broken only softly.
As interesting results, we will see that not only the Yukawa 
couplings but also some of the soft SUSY breaking parameters are restricted 
just by the RG invariant relations. Thus the compositeness requirement
can reduce the number of the free parameters in the supersymmetric theories.
In the next section we first consider the rigid cases. The SUSY breaking terms
are considered in Section 3. Section 4 is devoted to conclusions.

\section{Supersymmetric models with composite Higgs}

We consider the asymptotically free supersymmetric $SU(N_c)$ gauge
theory with $N_f (N_f < 3N_c)$ flavors of "quarks" 
$(Q_i^a, \tilde{Q}^i_a)$ $(a=1,\cdots,N_c, i=1,\cdots,N_f)$
belonging to the fundamental representation. We also introduce
the "Higgs" chiral fields $H_i^j$ in $(N_f, \bar{N_f})$ 
representation of the flavor $U(N_f)\times U(N_f)$ symmetry.
The Lagrangian with Yukawa interactions among quarks and Higgs is 
given by 
\bea
{\cal L}_{\rm gHY} &=&
\int d^4 \theta ( Q^{i \dagger} e^V Q_i + 
\tilde{Q}^i e^V \tilde{Q}_i^{\dagger}) 
+ \frac{1}{16 g^2} \int d^2 \theta W^A W_A + \mbox{h.c.} \nn \\
& &+\int d^4 \theta H^{\dagger} H 
+ \int d^2 \theta (y \tilde{Q}^j H_j^i Q_i  
+ \frac{m}{2} H_j^i H_i^j) + \mbox{h.c.}.
\label{ghyrigid}
\eea
Here we have assumed that the Yukawa coupling $y$ and the Higgs mass 
parameter $m$ are flavor independent just for the sake of
simplicity.
Note that the flavor $U(N_f)\times U(N_f)$  symmetry is 
broken down into the diagonal one $U(N_f)$ by 
the last mass term.
If the running Yukawa coupling $y(\mu)$ satisfies the so-called
compositeness condition at a certain scale $\Lambda$ (compositeness
scale):
\be
\lim_{\mu \rightarrow \Lambda} \frac{1}{y^2(\mu)} = 0,
\label{condy}
\ee
then the Higgs fields are reduced to be composite auxiliary fields
$H_i^j \sim Q_i \tilde{Q}^j $ at $\Lambda$.
After integrating out these fields we obtain the gauge theory
with dimension 5 interactions:
\bea
{\cal L}_{\rm D5} 
&=&\int d^4 \theta ( Q^{i \dagger} e^V Q_i 
+ \tilde{Q}^i e^V \tilde{Q}_i^{\dagger}) 
+ \frac{1}{16 g^2} \int d^2 \theta W^A W_A + \mbox{h.c.} \nn \\
& &- \int d^2 \theta \frac{h}{2} (Q_j \tilde{Q}^i) (Q_i \tilde{Q}^j) 
 + \mbox{h.c.},
\label{d5rigid}
\eea
where the coupling $h$ is given by
\be
\lim_{\mu \rightarrow \Lambda} \frac{y^2(\mu)}{m(\mu)} = h .
\ee
In the followings we examine the 1-loop RG equations for the 
gauge-Higgs-Yukawa theory given by Eq.~(\ref{ghyrigid}) and the solutions
satisfying the compositeness condition (\ref{condy}).

The beta functions for the N=1 SUSY gauge theories have been
known to be given exactly in terms of the anomalous dimensions of
the chiral matter fields. For our example they are found to be
\bea
& &\beta_g = \mu \frac{dg}{d\mu} =
-\frac{g^3}{16\pi^2} \frac{3N_c-N_f + N_f\gamma_Q}{1-(N_c/8\pi^2)g^2} \nn \\
& &\beta_y = \mu \frac{dy}{d\mu} =
(2\gamma_Q + \gamma_H)y \nn \\
& &\beta_m = \mu \frac{dm}{d\mu} =
2\gamma_H m,
\label{exactbeta}
\eea
where $\gamma_Q~(= \gamma_{\tilde{Q}})$ and $\gamma_H$ are the 
anomalous dimensions of the chiral fields $Q~(\tilde{Q})$ and 
$H$ respectively.
The wave function renormalization factors are given by 
\be
\gamma_Q = \frac{1}{2} \frac{d \ln Z_Q}{d \ln \mu},~~~
\gamma_H = \frac{1}{2} \frac{d \ln Z_H}{d \ln \mu}.
\ee
In the 1-loop perturbation the anomalous dimensions are given explicitly
as follows:
\be
\gamma_Q = \frac{1}{16\pi^2} (-\frac{N_c^2-1}{N_c} g^2 + N_f y\bar{y}),~~~
\gamma_H = \frac{N_c}{16\pi^2} y\bar{y}.
\label{anomalousdim}
\ee
Therefore we obtain the 1-loop beta functions for $\ag=g^2/8\pi^2$,
$\ay=y\bar{y}/8\pi^2$ and $m$ as
\bea
& & \frac{d \ag}{dt} = -b \ag^2, \label{betag} \\
& & \frac{d \ay}{dt} = (a \ay - c \ag) \ay, \label{betay} \\
& & \frac{d m}{dt} = d \ay m,\label{betam}
\eea
where we have introduced new parameters; $t=\ln(\mu/\mu_0)$, $a=2N_f + N_c$, 
$b=3N_c-N_f$, $c=2(N_c^2-1)/N_c$ and $d=N_c$.

The general solutions of Eq.~(\ref{betag}) and (\ref{betay}) 
have been analyzed in Ref.~\cite{pr,hkkn}. 
The solutions of Eq.~(\ref{betag}) are given by
\be
\ag(t) = \frac{\ag(0)}{1 + b\ag(0) t}.
\ee
By noting that the following quantity
\be
\Gamma = 
\ag^{\frac{c-b}{b}}\left(1 - \frac{c-b}{a} \frac{\ag}{\ay} \right)
\label{rgiy}
\ee
gives a RG invariant, the general solutions of Eq.~(\ref{betay}) are easily found 
to be
\be
\ay(t) = \frac{c-b}{a}\ag(t) \left( 1 + h_0 \ag^{-\frac{c-b}{b}}(t) \right)^{-1},
\label{soly}
\ee
where $h_0$ is an integration constant. The special solution with $h_0=0$,
$\ay^*(t)= (c-b)/b \ag(t)$, 
is known as the Pendleton-Ross "fixed point" \cite{pr}.
In the case of $c > b$, the Yukawa interaction is nontrivial if and only
if $h_0 \geq 0$. Therefore if the theory is nontrivial, then the Yukawa 
coupling at the IR scale $\mu_0$ should be observed in the region
\cite{kubo,hkkn}
\be
0 < \ay(0) \leq \ay^*(0) = \frac{c-b}{a} \ag(0).
\ee
While, for $-\ag^{(c-b)/b}(0) < h_0 < 0$
\footnote{We should exclude $h_0 < -\ag^{(c-b)/b}(0)$ so as for $\ay(0)$ to
be finite and positive.
}
, there exists a 
Landau pole at a finite value of $t >0$. Thus the solution satisfying 
the compositeness condition (\ref{condy}) at $t=\tl=\ln(\Lambda/\mu_0)$ 
is found to be
\be 
\ay(t;\tl) = \frac{c-b}{a}\ag(t) 
\left(1 - \left(\frac{\ag(t)}{\ag(\tl)}\right)^{-\frac{c-b}{b}} \right)^{-1}.
\ee
The Yukawa coupling of the gauge-Higgs-Yukawa theory equivalent to the
dimension 5 interaction model given at the scale $\Lambda$ is fixed to this
value. It should be noted that the sequence of theories parameterized by
$\Lambda$ converges to a nontrivial theory in the $\Lambda \rightarrow \infty$
limit.
This implies the (nonperturbative) renormalizability of the dimension 5
interaction models, since the models possess nontrivial continuum limits,
though perturbatively nonrenormalizable.
The Yukawa coupling corresponding to these renormalizable models is given by
the "fixed point" value $\ay^*(t)$. Such a nontrivial limit is realized 
only when $c > b, (N_c + 2/N_c < N_f < 3N_c)$. 
Hereafter we restrict ourselves to this case. 

Here let us sketch how the quantum corrections are treated in order to 
make the dimension 5 interaction models renormalizable.
The nonrenormalization theorem for supersymmetric theories holds
even in the presence of nonrenormalizable couplings \cite{weinberg}.
Therefore the superpotential is protected from quantum corrections
and we need to consider only the K\"{a}hler potential
part of $Q$ and $\tilde{Q}$.
In the 2-loop calculation the dimension 5 interaction causes
the quadratic divergence in the wave function renormalization factor 
$Z_Q$ \cite{fl}. 
Similarly higher powers of divergences are generated through the
higher loop corrections. This is the reason of the nonrenormalizability
in the perturbative calculations. 
However if we sum up the chain diagrams shown in Fig.~1 first, then
the four point function is given by the tree diagram of the composite 
Higgs exchange
\footnote{In the leading order of large $N_c$ expansion, the four point
function is evaluated in this way.}
. 
As a result, the power of divergence in the corrections for the two point
function $\langle Q Q^{\dagger} \rangle$ as shown in Fig.~2 is seen to be 
tamed down to logarithmic one.
Of course we do not say that the renormalizability proof is given by this
naive argument.
However the nonperturbative calculations have been explicitly performed for
the gauged NJL models by using the Exact RG, 
which is a sort of the Wilson RG \cite{kt}. 
In practice it has been found that the renormalized trajectory exists
as long as $c>b$. This implies the nonperturbative renormalizability
of the models. Here, however, we are not going to address to such
a nonperturbative analysis for the supersymmetric theories.

\begin{figure}[htb]
\epsfxsize=0.7\textwidth
\begin{center}
\leavevmode
\epsfbox{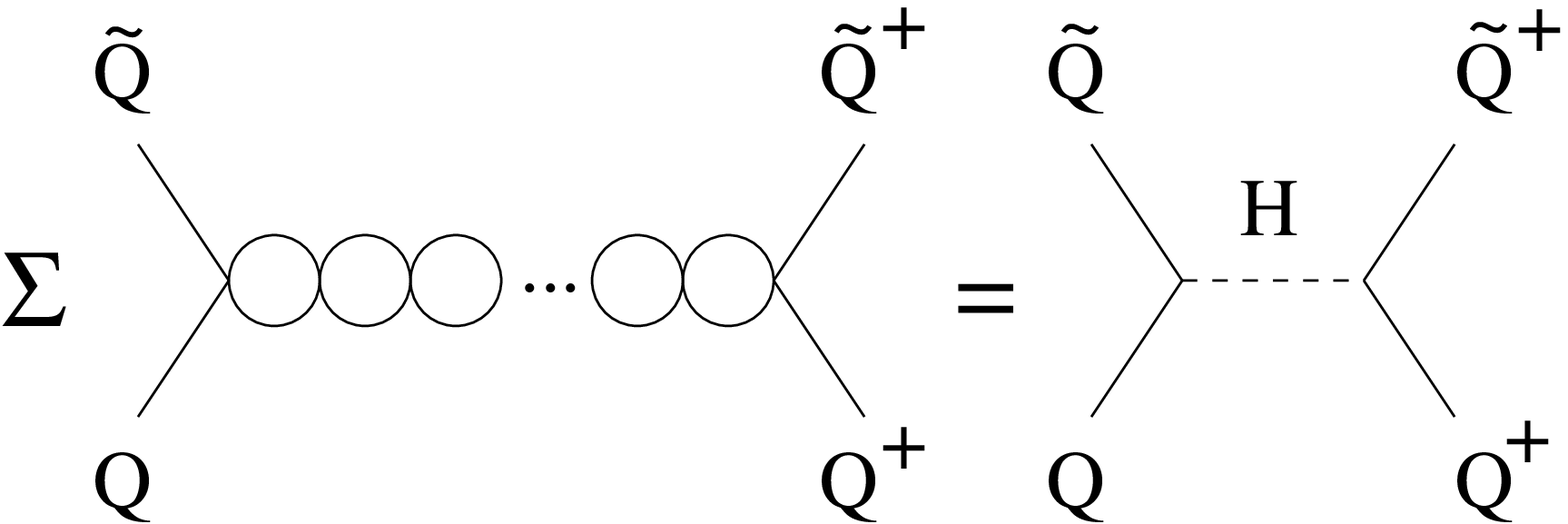}
\end{center}
\caption{The nonperturbative sum of the chain diagrams for the four point 
function.}
\epsfxsize=0.5\textwidth
\begin{center}
\leavevmode
\epsfbox{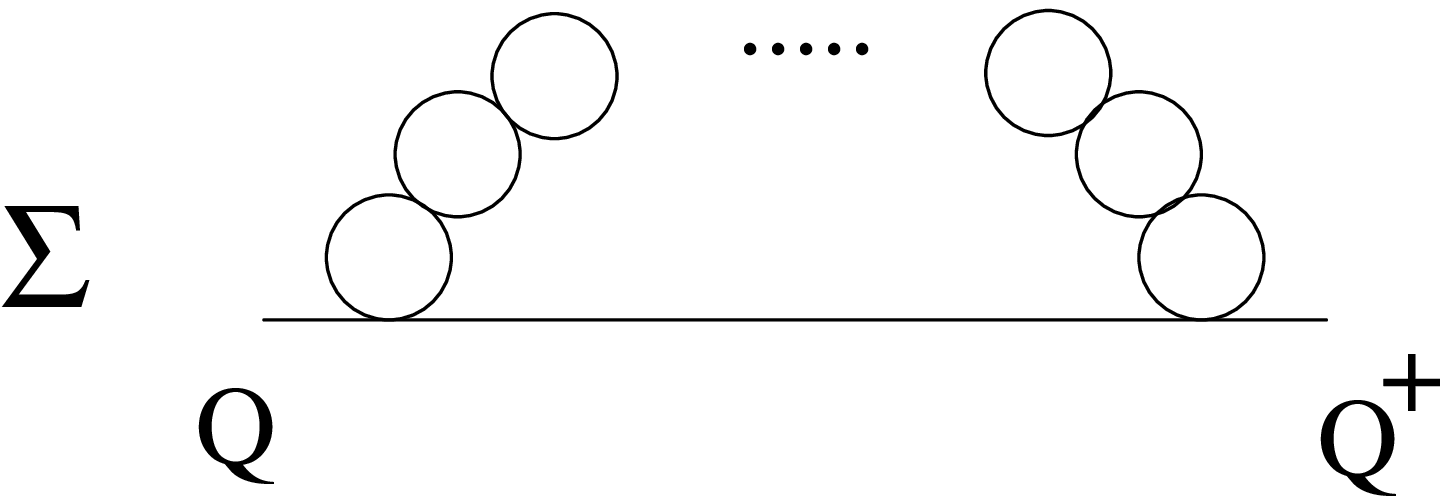}
\end{center}
\caption{The nonperturbative corrections to the two point function.}
\end{figure}

In practice the nontriviality of the Yukawa coupling is not sufficient
for the proof of renormalizability.
It is necessary to show that the supersymmetric mass parameter
$m$ also can be kept finite in the limit of 
$\Lambda \rightarrow \infty$.
If we can make the mass finite by tuning the coupling $h$ properly, 
then it will be shown that the dimension 5 interaction model is equivalent
to the nontrivial gauge-Higgs-Yukawa theory, therefore is 
renormalizable.
In order to see this, we first consider the renormalization of 
the mass parameter $m$ in the gauge-Higgs-Yukawa theory. 
The general solution of Eq.~(\ref{betam})
is also found from the RG invariant
\be
\Omega=m\left( \ay \ag^{-\frac{c}{b}} \right)^{-\frac{d}{a}}.
\label{rgim}
\ee
Therefore the mass parameter at scale $t$ is given by
\be
m(t)= m(0) \left(\frac{\ag(t)}{\ag(0)} \right)^{-\frac{cd}{ab}}
\left(\frac{\ay(t)}{\ay(0)} \right)^{\frac{d}{a}}.
\ee
It is seen that $m(t)$ also diverges as $t \rightarrow \tl$.

The parameter of the dimension 5 interaction at the compositeness
scale is given by
\be
|h|^2 = \frac{1}{\Lambda^2} |\hat{h}(\tl)|^2
= (8\pi^2)^2 \lim_{t \rightarrow \tl}\frac{\ay^2(t)}{|m(t)|^2},
\ee
where we have introduced the dimensionless parameter $\hat{h}$.
Therefore the mass parameter is related with the bare
coupling $\hat{h}(\tl)$ by
\be
|m(t)|^2 = (8\pi^2)^2 \ay^2(t)
\left(\frac{\ay(\tl)}{\ay(t)}\right)^{2-\frac{2d}{a}}
\left(\frac{\ag(\tl)}{\ag(t)}\right)^{\frac{2cd}{ab}}
\frac{\Lambda^2}{|\hat{h}(\tl)|^2}.
\ee
Indeed this expression is rather formal, since $\ay(\tl)$ is 
diverging.
However we can tune the bare coupling $\hat{h}(\tl)$ also diverging
so that the mass parameter $m(t)$ is maintained to be finite.
This property is a contrast with the four fermi coupling in the 
gauged NJL model, which is fine tuned to a finite
critical value \cite{hkkn,kt}.
Also the $\Lambda$ dependence of $m(t)$ is removed by adjusting
the bare coupling $\hat{h}(\tl)$ properly. 
The renormalization of $\hat{h}(\tl)$ is performed as 
\be
\frac{\mu}{\hat{h}(t)} = 
\left(\frac{\ay(\tl)}{\ay(t)}\right)^{1-\frac{d}{a}}
\left(\frac{\ag(\tl)}{\ag(t)}\right)^{\frac{cd}{ab}}
\frac{\Lambda}{\hat{h}(\tl)}.
\ee
With this renormalized coupling the mass parameter is expressed
by
\be
|m(t)|^2 = (8\pi^2)^2 \mu^2 \frac{\ay^2(t)}{|\hat{h}(t)|^2}.
\ee

Thus we have shown that the dimension 5 interaction models are
nonperturbatively renormalizable and the equivalent gauge-Higgs-Yukawa
theories must satisfy the IR fixed point relation: 
$\ay(t)/\ag(t) = (c-b)/b$. Such reduction of couplings 
\cite{reduction,kubo}
seems to necessarily happen, since the dimension 5 interaction model 
contains less number of the free parameters than the gauge-Higgs-Yukawa
theory does. 

In the rest of this section let us make a brief comment on the
relation to the magnetic description of SQCD in the conformal
window, $(3/2)N_c < N_f < 3N_c$ \cite{seiberg}.
It has been known that the gauge-Higgs-Yukawa theories given by Eq.~(1)
possess a nontrivial IR fixed point
and the IR dynamics is equivalently described by the $SU(N_f-N_c)$ SQCD
with $N_f$ flavors, if $m=0$.
The IR fixed point appears in the weak coupling region for
$3N_c=(1+\epsilon)N_f (\epsilon \ll 1)$ and therefore perturbation is
reliable. By using the anomalous dimensions evaluated at the 1-loop order
given in Eq.~(\ref{anomalousdim}), 
we find the fixed points at
\be
\begin{array}{ccll}
N_c(\ag^*, \ay^*) & = & (0, 0) & \mbox{UV fixed point}, \\
& = & (0, \epsilon) & \mbox{Unstable fixed point}, \\
& = & (2\epsilon, 7\epsilon) & \mbox{IR fixed point},
\end{array}
\ee
where we have assumed $N_c, N_f \gg 1$ \cite{ksv}
\footnote{The RG flow diagram is described in Ref.~\cite{ksv}.}.
Then the straight line connecting the UV fixed point and the IR fixed point
is just the special solution satisfying $\ay(t)/\ag(t) = (c-b)/a$
\cite{oehme}.
Therefore the RG analysis performed in this section implies that the IR
dynamics of the dimension 5 interaction model is described by
the theory on the IR fixed point added the mass perturbation,
$\int d \theta^2 HH + {\rm h.c.}$.
On the other hand the fixed point theory has the dual description by
the $SU(N_f-N_c)$ gauge theory. 
It is, therefore, speculated that the IR dynamics of the dimension 
5 interaction model is also described by the dual gauge theory perturbed
with the corresponding interaction 
$\int d \theta^2 (q \tilde{q})^2 + {\rm h.c.}$,
where $(q, \tilde{q})$ are the dual quarks of $SU(N_f-N_c)$ gauge theory.
This operator is found to have the same dimension as $\int d \theta^2 HH$
and, therefore, is relevant at the IR fixed point.

As $N_f$ reduces, the nontrivial fixed points moves towards the
strongly coupled region. In such cases perturbative calculation
is not applicable. However we can deduce some results from the
exact RG equations given by Eq.~(\ref{exactbeta}).
First it is seen that the anomalous dimensions at the IR fixed point 
are exactly obtained as
\be
\gamma_Q^* = -\frac{1}{2}\gamma_H^* = -\frac{3N_c-N_f}{N_f},
\ee
and that the Yukawa coupling becomes exactly marginal there.
Therefore the mass parameter $m$, whose dimension is given by 
$1 - \gamma_H$, is relevant for $N_f > 2N_c$. However it turns
irrelevant for $N_f < 2N_c$ and the composite Higgs particles $H$
become massless in the IR.
On the other hand the coupling of the dimension 5 interaction $h$
carries the dimension of $-1-2\gamma_Q$. Hence $h$ appears as a
relevant coupling for $N_f < 2N_c$, while it is irrelevant for
$N_f > 2N_c$.
Taking into account that the flavor symmetry $U(N_f)\times U(N_f)$ of
the gauge-Higgs-Yukawa theory is broken to diagonal $U(N_f)$
by the mass term, we should suppose that the operator
$\int d^2\theta (Q \tilde{Q})^2$
is generated in the IR by nonperturbative dynamics,
since it is the relevant operator allowed by the symmetry.
Then the IR fixed point plays a role of the UV fixed point for
the relevant perturbation by $h$. 
In this case the renormalizability of the theory with the 
"nonrenormalizable" interaction $\int d^2\theta (Q \tilde{Q})^2$ 
is guaranteed by the presence of the nontrivial UV fixed point.

\section{Soft SUSY breaking}

The spurion technique is very useful to incorporate the soft 
SUSY breaking parameters \cite{gg}.
Also it has been known that the beta functions for the soft
breaking parameters can be obtained immediately from the beta 
functions for the rigid theories \cite{yamada,jjp,softbeta,kazakov}.
For the gauge-Higgs-Yukawa theory examined in the previous section
we can introduce five soft SUSY breaking parameters 
$(m_Q^2, m_H^2, m_g, A, B)$. The Lagrangian is given as
\bea
{\cal L}_{\rm gHY}
&=&\int d^4 \theta 
(1 - m_Q^2 \eta \etab )
( Q^{i \dagger} e^V Q_i + \tilde{Q}^i e^V \tilde{Q}_i^{\dagger}) 
+ \int d^4 \theta 
(1 - m_H^2 \eta \etab ) H^{\dagger} H \nn \\
& & + \frac{1}{16g^2} \int d^2 \theta 
(1 - 2 m_g \eta) W^A W_A + \mbox{h.c.} \nn \\
& & + \int d^2 \theta 
\left( y(1 - A \eta) H_j^i Q_i \tilde{Q}^j 
+ \frac{m}{2} (1 - B \eta) H_j^i H_i^j
\right) + \mbox{h.c.},
\label{ghybroken}
\eea
by using the spurion superfields $\eta=\theta^2, \etab=\bar{\theta}^2$.
We also define the chiral superfields 
\bea
S &=& \frac{1}{g^2}(1 - 2m_g \eta), \nn \\
Y &=& y(1-A \eta), \nn \\
M &=& m(1 - B \eta),
\label{d5broken}
\eea
for the later conveniences.
On the other hand we suppose to introduce the supersymmetry breaking 
parameters in the dimension 5 interaction model as 
\bea
{\cal L}_{\rm D5}
&=&\int d^4 \theta 
(1 - m_Q^2 \eta \etab )
( Q^{i \dagger} e^V Q_i + \tilde{Q}^i e^V \tilde{Q}_i^{\dagger}) \nn \\
& & + \frac{1}{16 g^2} \int d^2 \theta 
(1 - 2 m_g \eta) W^A W_A 
- \int d^2 \theta 
\frac{h}{2}(1 - C \eta)(Q_j \tilde{Q}^i)(Q_i \tilde{Q}^j) 
+ \mbox{h.c.}.
\eea

In this section we are going to examine whether this model can be
also equivalent to the specific case of the softly broken 
gauge-Higgs-Yukawa theories. 
Note that the $C$-term introduced into the Lagrangian (\ref{d5broken}) 
gives a SUSY breaking term of four-scalar interaction. In the case of 
perturbatively renormalizable theories such a term is regarded as hard
breaking.
However, our findings are as follows. We can show the equivalence 
between these
two types of theories by imposing certain compositeness conditions
just as seen in the rigid cases.
Therefore it is seen that the dimension 5 interaction models are
actually softly broken as well as renormalizable by taking the 
infinite limit of the compositeness scale.
Moreover it is found that the soft SUSY breaking parameters
$(m_Q^2, m_H^2, A)$ in the corresponding gauge-Higgs-Yukawa theories
are restricted according to the RG invariant relations 
\cite{jjp,jj1,kkk}.

The beta functions for the soft SUSY breaking parameters can be obtained
by introducing the renormalization superfield $\tilde{Z_i} (i=V, Q, H)$ 
related to $Z_i$ via the redefinition of coupling constants 
\cite{yamada,kazakov,jjp}
\footnote{
The structure of the superfield diagrams restricts the Yukawa coupling 
dependence of the $Z$ factor so that 
$\beta_y (\partial Z/\partial y)=\beta_{\bar{y}}
(\partial Z/\partial \bar{y})$ \cite{jj1}.
Therefore $Z$ is given in terms of $\ay$ in the rigid case.
}
\be
\tilde{Z}_i (\ag, \ay) = Z_i (\tilde{\ag}, \tilde{\ay}).
\label{zfactor}
\ee
Here the redefined couplings are given by
\bea
\tilde{\ag} &=& \frac{1}{8\pi^2} \left[{\rm Re}(S) \right]^{-1} \\
&=& \ag(1 + m_g \eta + \bar{m}_g \etab + 2m_g\bar{m}_g\eta\etab ), \nn \\
\tilde{\ay} &=& \frac{1}{8\pi^2}
Y \bar{Y}(1 + m_Q^2\eta \etab)^2(1+m_H^2\eta\etab ) \\
&=& \ay(1 - A \eta - \bar{A} \etab + (A\bar{A} + \Sigma)\eta \etab), 
\eea
where $\Sigma = 2m_Q^2 + m_H^2$.
By using this fact and the anomalous dimensions given by 
Eq.~(\ref{anomalousdim}),
the 1-loop beta functions for the couplings $(m_Q^2, m_H^2, m_g, A, B)$ 
\footnote{The exact beta functions of the soft scalar mass parameters
contain an extra regularization scheme dependent term, which is not
obtained by this procedure. (See Ref.~\cite{jjp}.) 
However this term does not contribute in the 1-loop order analysis 
performed here. 
}
are found to be 
\bea
\frac{d m_g}{dt}&=&-b \ag m_g, \label{betamg} \\
\frac{d A}{dt}&=&a \ay A + c \ag m_g, \label{bataA} \\
\frac{d B}{dt}&=&d \ay A, \label{bataB} \\
\frac{d m_Q^2}{dt}&=&-c m_g\bar{m_g} \ag + N_f(A\bar{A} + \Sigma) \ay, 
\label{batamQ} \\
\frac{d m_H^2}{dt}&=&N_c(A\bar{A} + \Sigma) \ay. \label{betamH}
\eea
Therefore the sum of the soft scalar mass $\Sigma$ satisfies
\be
\frac{d \Sigma}{dt} = - 2c \ag m_g\bar{m}_g + a\ay (A\bar{A} + \Sigma).
\label{betasum}
\ee

Now we seek for the general solutions of these RG equations in order to
extract the specific RG flows allowed by the compositeness conditions.
In the softly broken case also, it is efficient to find out
the RG invariants to obtain the general solutions \cite{ky} \footnote{
See also Ref.\cite{kazakov2}.}.
Interestingly enough the renormalizations of the combinations
${\rm Re}(S)$, $\tilde{\ay}$ and 
$\tilde{|m|}^2 = M\bar{M}(1 + m_H^2\eta \etab )^2$ are performed 
with the renormalization factors given by the redefinition of 
coupling constants;
\bea
{\rm Re}(S)_{\rm bare}&=&Z_V^{-1}(\tilde{\ag},\tilde{\ay})
{\rm Re}(S)_{\rm ren}, \label{reng} \\
\tilde{\ay}_{\rm bare}&=&Z_Q^{-2}(\tilde{\ag},\tilde{\ay})
Z_H^{-1}(\tilde{\ag},\tilde{\ay}) \tilde{\ay}_{\rm ren}, 
\label{reny} \\
\tilde{|m|}^2_{\rm bare}&=&Z_H^{-2}(\tilde{\ag},\tilde{\ay})
\tilde{|m|}^2_{\rm ren}. \label{renm}
\eea
It should be noted that these are just the same forms of the 
renormalization in the rigid case and hold in all orders
of perturbation. Here let us mention the reason why these
renormalizations are satisfied. Eq.~(\ref{reng}) follows from
the renormalization of the vector superfield $V$ naively.
On the other hand the wave function renormalization factors of 
$Q, \tilde{Q}$ and $H$ must be chiral superfields.
Therefore the renormalization superfields 
$\tilde{Z_i} (i= Q, H)$ given by 
Eq.~(\ref{zfactor}) should be decomposed into the chiral
(antichiral) renormalization superfields $z_i (\bar{z}_i)$
and the renormalization of soft scalar masses $\Delta_i$ as
\be
\tilde{Z}_i = z_i (1 + \Delta_i \eta \etab ) \bar{z}_i.
\ee
The renormalization of the chiral fields $Y, M$ is
given in terms of $z_i$ \cite{kazakov} by
\bea
Y_{\rm bare} &=& z_Q^{-2}z_H^{-1} Y_{\rm ren}, \nn \\
M_{\rm bare} &=& z_H^{-2} M_{\rm ren}.
\eea
The renormalizations for the antichiral superfields are
given similarly.
The renormalization for the soft scalar masses is also 
performed by
\be
(1 + m^2_i \eta \etab )_{\rm bare} 
= (1 + \Delta_i \eta \etab )^{-1}(1 + m^2_i \eta \etab )_{\rm ren}. 
\ee
By taking account of these renormalizations, Eq.~(\ref{reny}) and
(\ref{renm}) are found to realize.

Therefore the 1-loop beta functions for the above combinations 
can be immediately written down as
\bea
\frac{d}{dt} \tilde{\ag}^{-1}&=&-b, \label{betaredefg} \\
\frac{d}{dt} \tilde{\ay}&=& 
(a\tilde{\ay}-c\tilde{\ag})\tilde{\ay},\label{betaredefy} \\
\frac{d}{dt} \tilde{|m|}^2&=&
2d \tilde{\ay} \tilde{|m|}^2. \label{betaredefm}
\eea
Certainly these equations are found to reproduce 
Eq.~(\ref{betamg}-\ref{betasum}). 
The RG invariants for the redefined couplings are
also easily obtained, since we have already known the RG 
invariants in the rigid case as given by 
Eq.~(\ref{rgiy}) and (\ref{rgim}).

First let us consider the RG invariant derived from 
Eq.~(\ref{betaredefg}).
It is seen that the $\eta$ ($\etab$) component of the L.H.S. gives
a RG invariant $m_g/\ag$ ($\bar{m}_g/\ag$) \cite{hs,jjp,jj1}.
Therefore the general solution of the gaugino mass is found to be
\be
m_g(t) = m_g(0)\left( \frac{\ag(t)}{\ag(0)}\right).
\ee
Next consider the RG invariant for $\tilde{\ay}$ \cite{ky}.
The RG invariant $\Gamma$ given by Eq.~(\ref{rgiy}) is extended to
\be
\tilde{\ag}^{\frac{c-b}{b}}
\left(1 - \frac{c-b}{a} \frac{\tilde{\ag}}{\tilde{\ay}} \right)
= \tilde{\Gamma} = -h_0 -h'_0 \eta- \bar{h'}_0 \etab 
+ h''_0 \eta\etab,
\ee
which generates a set of RG invariants.
These RG invariant quantities present us the general solutions for 
$A$ and $\Sigma$ adding to Eq.~(\ref{soly}):
\bea
A(t) &=& -m_g(t) + \frac{c-b}{b}(R(t)-1)m_g(t) 
+ h'_0 \ag^{-\frac{c-b}{b}}(t)R(t), \\
\Sigma(t) &=& |m_g(t)|^2 + 
(A(t)+m_g(t)) (\bar{A}(t) + \bar{m}_g(t) ) \nn \\
& &+\frac{c-b}{b}\left[
(A(t)+m_g(t) )\bar{m}_g(t)
+ (\bar{A}(t)+\bar{m}_g(t) ) m_g(t)
\right] \nn \\
& &-\frac{c}{b}\frac{c-b}{b}(R(t)-1)|m_g(t)|^2
+h''_0 \ag^{-\frac{c-b}{b}}(t)R(t),
\eea
where we have introduced 
\be
R(t) = \frac{a}{c-b} \frac{\ay(t)}{\ag(t)}.
\ee
Similarly we derive a set of RG invariants by extending 
Eq.~(\ref{rgim}).
The RG invariant
\be
\tilde{|m|}^2 
\left(\tilde{\ay} \tilde{\ag}^{-\frac{c}{b}}\right)^{-\frac{2d}{a}}
= |\tilde{\Omega}|^2 = 
\exp (k_0 - k'_0\eta - \bar{k}'_0\etab + 2k''_0\eta \etab),
\ee
is found to present us the general solutions for 
$B$ and $m_H^2$ as
\bea
B(t)&=& \frac{2d}{a} \left( A(t) + \frac{c}{b}m_g(t) \right) + k'_0, \\
m_H^2(t) &=& \frac{d}{a} \left( \Sigma(t)-\frac{c}{b}|m_g(t)|^2 \right)
+k''_0.
\eea
Thus we have obtained the general solutions for all the running 
coupling constants appearing in the softly broken gauge-Higgs-Yukawa
theories.

Now let us consider the compositeness conditions 
for the soft parameters, i.e. 
\be
\lim_{\mu \rightarrow \Lambda} \frac{1}{\tilde \alpha_y(\mu)} = 0.
\ee
In order that the gauge-Higgs-Yukawa theory becomes equivalent to the 
model given by Eq.~(\ref{d5broken}) at the compositeness scale $\Lambda$, 
it is sufficient for all of $A(t), B(t), m_Q^2(t), m_H^2(t)$ to be
finite in the limit $t \rightarrow \tl$.
In the expressions of the general solutions 
$R(t)$ diverges in this limit. 
Therefore it is enough to fix the integration constants as
\bea
h'_0 &=& -\frac{c-b}{b}\ag^{\frac{c-b}{b}}(\tl)m_g(\tl), \nn \\
h''_0 &=& \frac{c}{b}\frac{c-b}{b}\ag^{\frac{c-b}{b}}(\tl)|m_g(\tl)|^2.
\eea
As a result $A(t)$ and $\Sigma(t)$ are completely fixed, while $B(t)$ and
$m_H^2(t)$ remain to be adjustable parameters. This result seems reasonable,
since the number of the independent couplings should be equal to
that of the dimension 5 interaction model.

If we take the limit $\Lambda \rightarrow \infty$, the dimension 5 interaction
models become equivalent to the nontrivial, therefore nonperturbatively 
renormalizable, and also softly broken gauge-Higgs-Yukawa theories.
In these theories the soft parameters $A$ and $\Sigma$ are reduced to
\bea
A(t) &=& -m_g(t), \nn \\
\Sigma(t) &=& |m_g(t)|^2,
\eea
which are the RG invariant relations \cite{jjp,jj1,kkk}.
Also the scale dependence of the running couplings $B(t)$ and $m^2_H(t)$ is
given by
\bea
B(t)&=& \frac{2d}{a}\frac{c-b}{b}m_g(t)+ k'_0, \nn \\
m_H^2(t) &=& -\frac{d}{a} \frac{c-b}{b}|m_g(t)|^2 + k''_0,
\eea
in this limit. It should be noted that these solutions coincide also 
with the RG invariant relations up to integration constants.

\section{Conclusions}
In this paper we have studied the special solutions of 
1-loop running couplings in
the softly broken supersymmetric gauge-Higgs-Yukawa theories 
satisfying the compositeness conditions.
These theories can be regarded as composite Higgs models and the
fundamental theories contain dimension 5 interactions. 
In the limit of infinite composite scale the theories become equivalent
to the gauge-Higgs-Yukawa theories with the nontrivial continuum
limits. Therefore the dimension 5 interaction models are found to be
renormalizable in this nonperturbative sense.

The models contain also a SUSY breaking term of four-scalar interaction,
which is regarded as hard breaking if added to the perturbatively 
renormalizable theories.
In the equivalent gauge-Higgs-Yukawa theories, however, supersymmetry is
only softly broken. 
It has been found that the A-parameter and the sum of soft masses $\Sigma$
as well as the Yukawa coupling of the composite Higgs models are completely
fixed by their RG invariant relations as the composite scale goes to infinity.
Thus the number of free parameters can be well reduced by considering
the composite models.
The solutions for the gaugino mass, the B-parameter and the soft masses
contain free parameters adjustable by the bare SUSY breaking parameters in
the fundamental theories.

The soft SUSY breaking parameters restricted to the RG invariants are also
obtained in the anomaly mediated SUSY breaking scenarios \cite{anomaly}.
However in these scenarios the slepton masses are predicted to have negative
(mass)$^2$, since each of the soft scalar masses is restricted by the RG 
invariant relation. 
Therefore, another source of SUSY breaking has to be introduced
in order to resolve this problem \cite{anomaly, dterm}.
In contrast with the anomaly mediated SUSY breaking, the composite Higgs
models allow us to adjust each soft scalar mass so that the RG invariant
sum rules are unchanged \cite{kkk}.
The soft scalar masses should be determined by the SUSY breaking 
mechanism in the fundamental models. 
However the composite Higgs model cannot restrict the couplings other 
than in the top sector, if applied to the MSSM. Therefore we cannot say
anything about the slepton masses in this approach.
Rather, we have presented this model as a possible dynamical mechanism 
to constrain the sum of soft scalar masses as well as the Yukawa
and the A-term couplings to the RG invariants \footnote{If 
even small Yukawa couplings are realized as fixed points, 
the sum rules are applicable to other sfermions as 
proposed in Ref.\cite{ky2}.}.

We have also discussed the IR dynamics of the gauge-Higgs-Yukawa model in 
relation to the Seiberg duality. In the conformal window, 
$(3/2)N_c < N_f < 3N_c$, the IR dynamics is controlled by the nontrivial
IR fixed point. Specially in the case of $N_f <2N_c$, the dimension 5
interaction $\int d^2\theta (Q\tilde{Q})^2$ are found to be relevant,
though the mass perturbation is irrelevant there.
The renormalizability of the theory deformed by this interaction is
guaranteed by the nontrivial fixed point.

\vspace*{3mm}
\noindent
{\bf \large Acknowledgments}

The authors would like to thank J.~Kubo and G.~Zoupanos for valuable 
discussions and comments.

\end{document}